\documentclass{kapproc}

\usepackage{procps} 
\usepackage[dvips]{graphicx}

\upperandlowercase
\kluwerbib

\begin{document}

\articletitle[]{SparsePak Observations of Diffuse Ionized Gas Halo Kinematics in NGC 891}

\chaptitlerunninghead{DIG Halo Kinematics in NGC 891}

\author{George H. Heald\altaffilmark{1}, Richard J. Rand\altaffilmark{1},
 Robert A. Benjamin\altaffilmark{2}, Matthew A. Bershady\altaffilmark{3}}

\affil{\altaffilmark{1}Univ. of New Mexico, \ 
\altaffilmark{2}Univ. of Wisconsin -- Whitewater, \
\altaffilmark{3}Univ. of Wisconsin -- Madison}

\begin{abstract}
We present WIYN SparsePak observations of the diffuse ionized gas (DIG) halo of NGC 891. Preliminary results of an analysis of the halo velocity field reveal a clear gradient of the azimuthal velocity with $z$ which agrees with results for the neutral gas. The magnitude of the gradient has been determined, using two independent methods, to be approximately 15 km s$^{-1}$ kpc$^{-1}$.
\end{abstract}

\section{Observations and Data Reduction}

\begin{figure}[ht]
\includegraphics[width=3in]{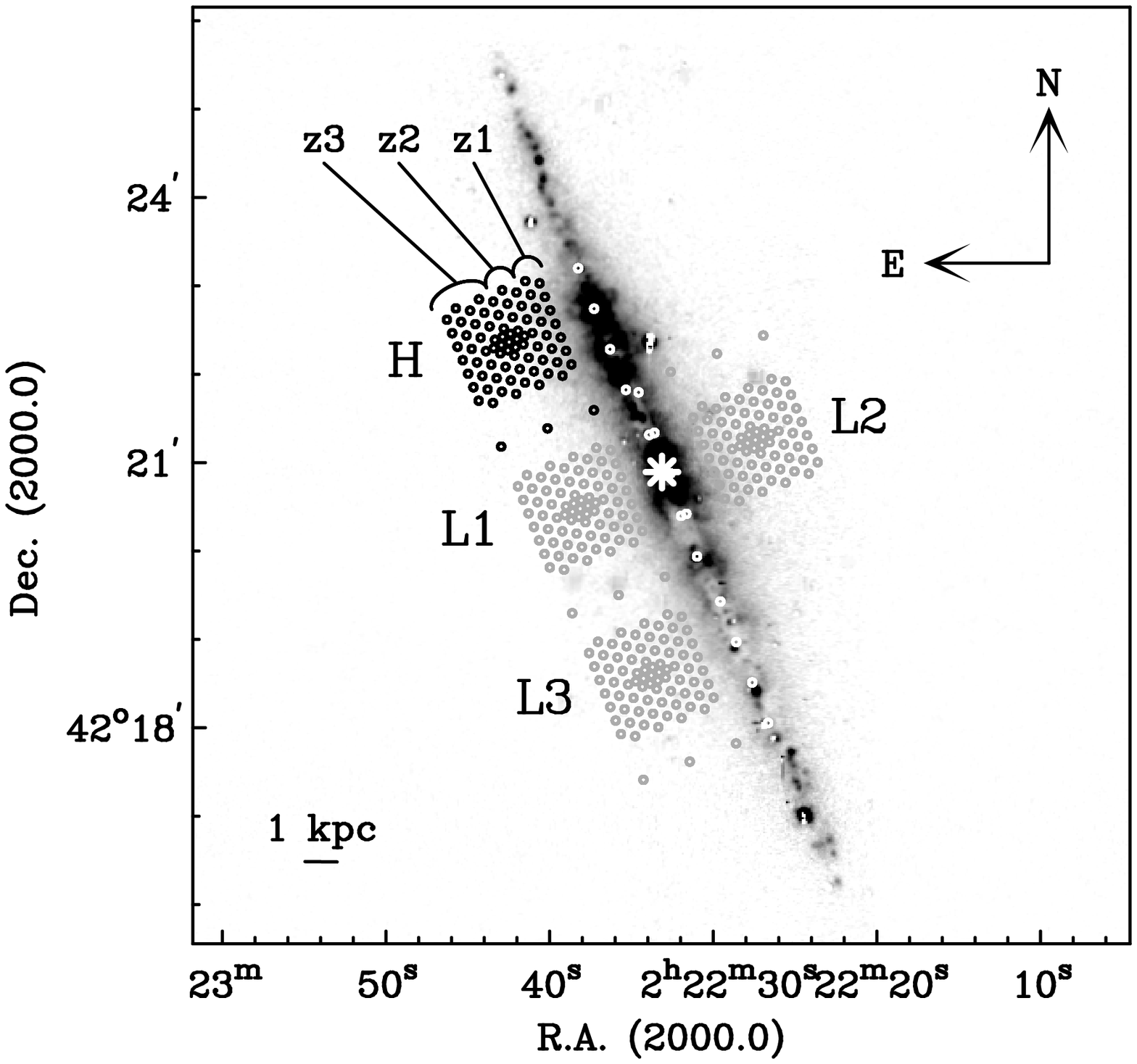}
\narrowcaption{H$\alpha$ image of NGC 891 from \cite{rand90}, with the positions of the SparsePak fibers during the higher spectral resolution observations (H) overlaid as black circles. Lower spectral resolution pointings (L1, L2, and L3), to be presented in a future paper, are plotted with gray circles. Fibers on the major axis are plotted in white for clarity. Ranges of $z$ used to construct PV diagrams (see text) are labeled z1, z2, and z3. The rotation center is marked with a white star. The receding side is to the south.}
\label{figure:image}
\end{figure}

Data were obtained during the nights of 10--12 December 2004 with the SparsePak IFU (see \cite{bershady04,bershady05}). We used the 860 l/mm grating at order 2, which provides a spectral resolution $\lambda/\Delta\lambda\sim$ 4,900, and the 316 l/mm grating at order 8, which provides a spectral resolution $\sim$ 10,000. In the latter mode, the wavelength coverage includes the H$\alpha$, [N$\,$II] $\lambda \lambda$6548,6583 and [S$\,$II] $\lambda \lambda$6716,6731 emission lines. The higher spectral resolution data are presented and discussed in this paper. The data were reduced in IRAF. Figure \ref{figure:image} shows the SparsePak pointings on an H$\alpha$ image of NGC 891.

\section{Azimuthal Velocities from Envelope Tracing}

To analyze the kinematic structure of the DIG halo of NGC 891, position-velocity (PV) diagrams have been constructed from the spectra of all 82 fibers in pointing `H' (cf. Fig. \ref{figure:image}). To enhance signal-to-noise, several fibers were averaged together for each major axis distance $R$ at the highest $z$. Figure \ref{figure:image} indicates which fibers were included in each $z$-range. Because a bright sky line interferes with several of the H$\alpha$ profiles, and to increase signal-to-noise, PV diagrams were constructed of the sum of the [N$\,$II] $\lambda$6583 + [S$\,$II] $\lambda$6716 emission lines. Contour plots of the PV diagrams are included in Figure \ref{figure:pvdiags}.

\begin{figure}[ht]
\hspace{0.5in}
\includegraphics[width=1.7in]{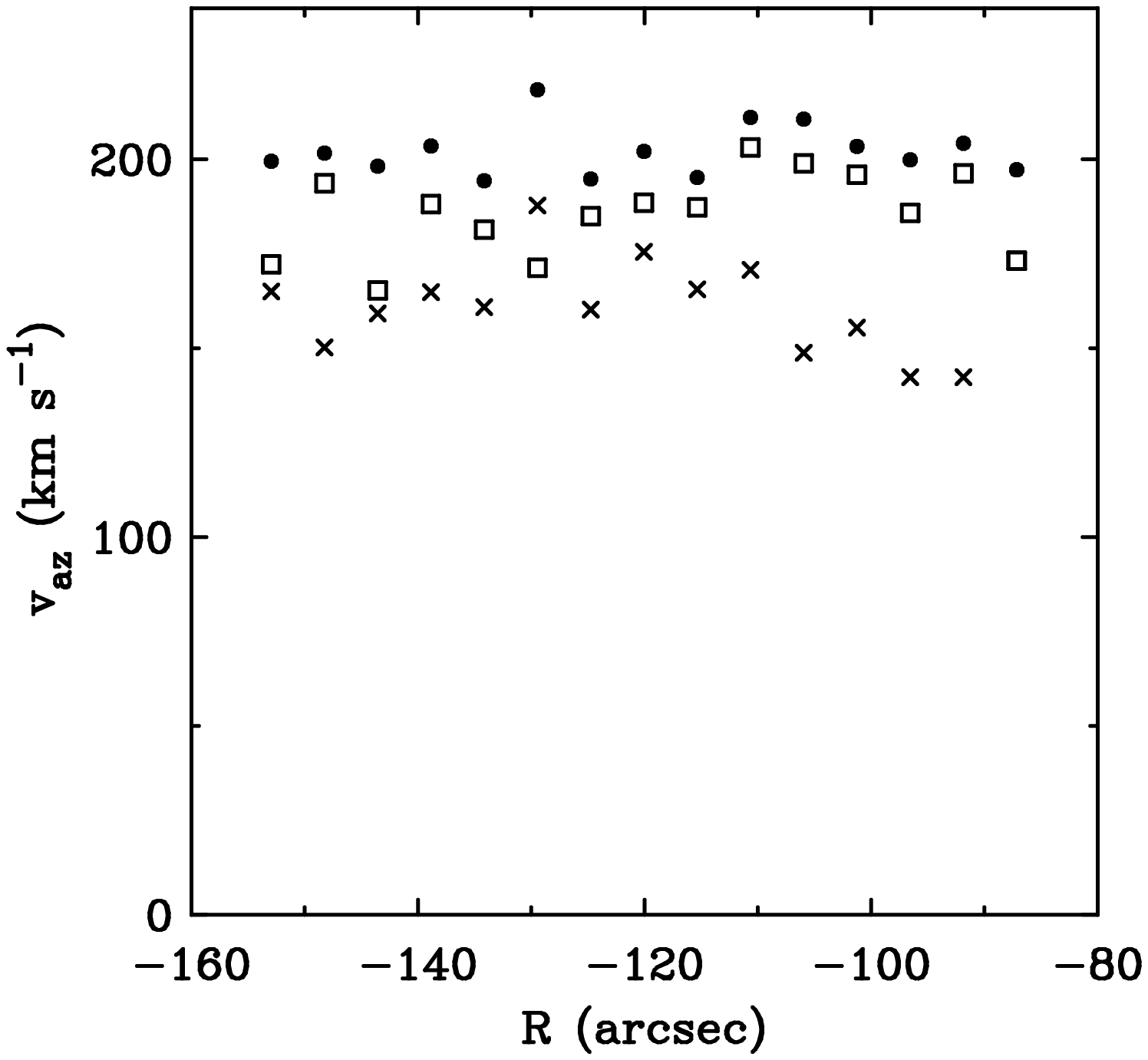}
\narrowcaption[width=3in]{Azimuthal velocity curves, derived from the PV diagrams shown in Figure 3 using the envelope tracing method. Velocities are shown for $25\mbox{$^{\prime\prime}$}<z<45\mbox{$^{\prime\prime}$}$ (z1; \emph{filled circles}), $45\mbox{$^{\prime\prime}$}<z<65\mbox{$^{\prime\prime}$}$ (z2; \emph{open squares}), and $z>65\mbox{$^{\prime\prime}$}$ (z3; \emph{crosses}), and are relative to $v_{\mathrm{sys}}=528$ km s$^{-1}$.}
\label{figure:rotcurs}
\end{figure}

Rotation curves were derived for each PV diagram using the envelope tracing method (e.g., \cite{sofue01}). Because NGC 891 is nearly edge-on ($i > 88\mbox{$^\circ$}$; \cite{swaters94}), the edge of each line profile furthest from the systemic velocity (the ``envelope'') corresponds to gas at the line of nodes; this velocity is thus approximately the azimuthal velocity at that $R$. The azimuthal velocity is found by (\cite{sofue01}):
\begin{equation}
v_{\mathrm{az}}=(v_{\mathrm{env}}-v_{\mathrm{sys}})/\sin(i)-\sqrt{\sigma_{\mathrm{inst}}^2+\sigma_{\mathrm{gas}}^2},
\end{equation}
where $v_{\mathrm{env}}$ is the velocity at the location of the envelope, $\sigma_{\mathrm{inst}}$ is the velocity resolution of the instrument, and $\sigma_{\mathrm{gas}}$ is the velocity dispersion of the gas. Azimuthal velocity curves were constructed from the PV diagrams, and are shown in Figure \ref{figure:rotcurs}. Extinction in the midplane prevented reliable velocities from being found for $z\approx 0\mbox{$^{\prime\prime}$}$.

It is clear from Figure \ref{figure:rotcurs} that the derived azimuthal velocities decrease with height. A close inspection of the PV diagrams shows that this is not an effect of signal-to-noise variations with height. An average azimuthal velocity gradient of 0.80 km s$^{-1}$ arcsec$^{-1}$ (17 km s$^{-1}$ kpc$^{-1}$) was calculated. Similar results were obtained by following this procedure for the [N$\,$II] $\lambda$6583 emission line alone (14 km s$^{-1}$ kpc$^{-1}$). These values are in agreement with that measured by \cite{fraternali04} for the neutral component, 15 km s$^{-1}$ kpc$^{-1}$.

\section{Azimuthal Velocities from PV Diagram Modeling}

The envelope tracing method is sensitive to the changing signal-to-noise ratio of the data with $z$. To account for this effect and the radial gas distribution, we have generated galaxy models. The H$\alpha$ image presented in Figure \ref{figure:image} was used to obtain estimates of the radial density profile at each of the heights considered in the model. Because the distribution of the DIG is not axisymmetric, the profiles were modified by hand to better match the shape of the data PV diagrams. The amplitude of the radial density profile was chosen such that the signal-to-noise in the model approximately matched that in the data.

\begin{figure}[ht]
\includegraphics[width=3.5in,angle=-90]{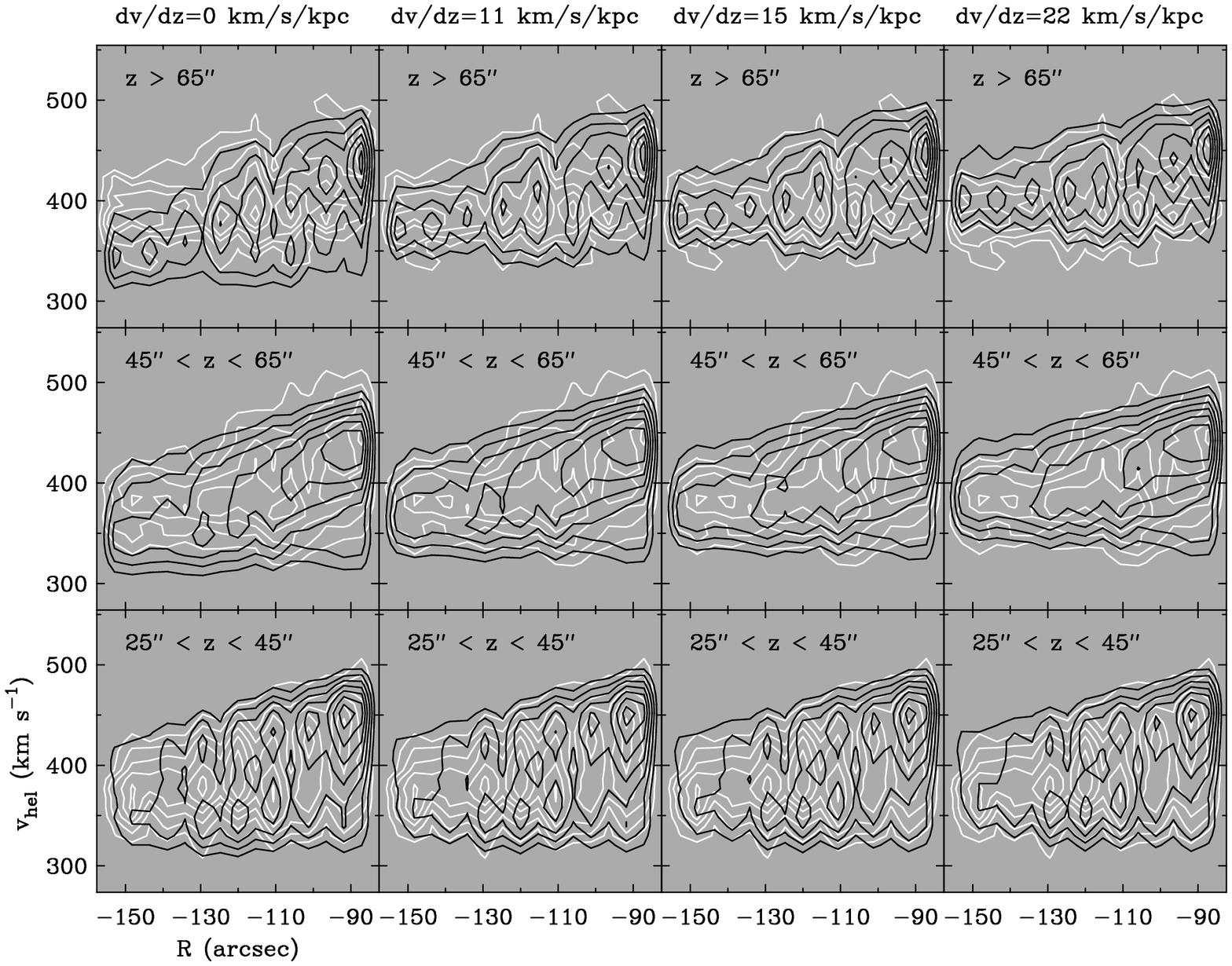}
\caption{Comparison between data (\emph{white}) and model (\emph{black}) PV diagrams for different values of $dv/dz$, at the indicated heights. Contour levels for data and models are 10, 15, 20, 25, 30, 35, and 40$\sigma$ for $25\mbox{$^{\prime\prime}$}<z<45\mbox{$^{\prime\prime}$}$ (z1); 5, 8, 11, 14, 17, and 20$\sigma$ for $45\mbox{$^{\prime\prime}$}<z<65\mbox{$^{\prime\prime}$}$ (z2); 3, 4.5, 6, 7.5, 9, 10.5, and 12$\sigma$ for $z>65\mbox{$^{\prime\prime}$}$ (z3). The systemic velocity is $v_{\mathrm{sys}}=528$ km s$^{-1}$.}
\label{figure:pvdiags}
\end{figure}

We then created model galaxies with these derived radial profiles using a version of the GIPSY task GALMOD, modified to allow for a vertical gradient in azimuthal velocity which begins at a height $z_0$ above the midplane:
\begin{equation}
v(R,z)=v(R,z\leq z_0)-\frac{dv}{dz}\,\left[|z|-z_0\right],
\end{equation}
where $dv/dz$ has units of [km s$^{-1}$ arcsec$^{-1}$], and for the models considered here, the major axis rotation curve is flat [$v(R,z\leq z_0)=200$ km s$^{-1}$]. The inclusion of the parameter $z_0$ was motivated by the results of \cite{fraternali04}, who find that the gradient in the neutral component of NGC 891 starts at approximately $z=1.3$ kpc (though that result may be a consequence of beam smearing). That value of $z_0$ was used for the models shown here, but we note that the appropriate value cannot be determined for our data (we lack data below $z\approx1.2$ kpc). Artificial SparsePak observations were made of the models to create PV diagrams, and compared to the data (Figure \ref{figure:pvdiags}).

The model that appears to best match the data has $dv/dz = 15$ km s$^{-1}$ kpc$^{-1}$. A statistical analysis of difference images suggests that the best match occurs at $dv/dz=12-14$ km s$^{-1}$ kpc$^{-1}$, but because the model cannot perfectly reproduce the shape of the PV diagrams, comparing by eye is the most reliable way to determine the gradient. To ensure that our results are not influenced by an incorrect specification of the radial density profile, we also consider models with a flat radial density profile, and find that the same gradient yields the best match. We conclude that the vertical gradient in azimuthal velocity for the DIG halo of NGC 891 is approximately 15 km s$^{-1}$ kpc$^{-1}$.

\section{Future Work}

To attempt to understand the origin of the azimuthal velocity gradient derived in this paper, the kinematic data will be compared with the results of an entirely ballistic model of disk--halo flow (cf. \cite{collins02}). Further information regarding the kinematic structure will be obtained once the lower spectral resolution pointings are reduced and analyzed. Results from these efforts will be presented in a forthcoming paper.

\begin{acknowledgments}
This material is based on work partially supported by the National Science Foundation under Grant No. AST 99-86113.
\end{acknowledgments}

\end{document}